\documentclass[oneside,article]{memoir}
\usepackage{amsmath}         
\usepackage{amsthm,thmtools,thm-restate} 
\usepackage{enumitem}        

\usepackage{tikz}
  \usetikzlibrary{matrix,arrows,positioning,calc}
  
\usepackage{tikz-cd}

\usepackage{float} 
\usepackage{caption, subcaption} 

\usepackage[nobysame,alphabetic,initials]{amsrefs} 
\usepackage[hidelinks]{hyperref}                   
\usepackage[capitalize,nosort]{cleveref}           

\usepackage{amssymb}
\usepackage[mathscr]{euscript}
\usepackage{relsize}
\usepackage{stmaryrd}
\usepackage{stackengine}

\usepackage{indentfirst} 
\usepackage{multicol}
\usepackage{algorithm}
\usepackage{algorithmic}
\usepackage[inline,nomargin]{fixme} 
  \fxsetup{targetlayout=color}

\usepackage{float} 

\isopage[12]

\setlrmargins{*}{*}{1}
\checkandfixthelayout

\counterwithout{section}{chapter}
\usepackage{appendix}

  \newcommand{\iso}{\mathrel{\cong}}

  \declaretheorem[style=definition,within=section]{definition}
  \declaretheorem[style=definition,numberlike=definition]{example}

  \declaretheorem[style=plain,numberlike=definition]{theorem}

  \declaretheorem[style=plain,numbered=no,name=Theorem]{theorem*}

  \Crefname{corollary}{Corollary}{Corollaries}
  \Crefname{definition}{Definition}{Definitions}
  \Crefname{lemma}{Lemma}{Lemmas}
  \Crefname{proposition}{Proposition}{Propositions}
  \Crefname{remark}{Remark}{Remarks}
  \Crefname{theorem}{Theorem}{Theorems}

  \newlist{axioms}{enumerate}{1}
  \Crefname{axiomsi}{}{}

  \newenvironment{tikzeq*}
  {
    \begingroup
    \begin{equation*}
    \begin{tikzpicture}[baseline=(current bounding box.center)]
  }
  {
    \end{tikzpicture}
    \end{equation*}
    \endgroup
    \ignorespacesafterend
  }

  \tikzset
  {
    diagram/.style=
    {
      matrix of math nodes,
      column sep={4.3em,between origins},
      row sep={4em,between origins},
      text height=1.5ex,
      text depth=.25ex
    },
    over/.style={preaction={draw=white,-,line width=6pt}},
    every to/.style={font=\footnotesize},
    inj/.style={right hook->},
    surj/.style={-{Latex[open]}},
    cof/.style={>->},
    fib/.style={->>},
  }

  \DeclareFontFamily{U}{mathx}{\hyphenchar\font45}

  \DeclareFontShape{U}{mathx}{m}{n}{
    <5> <6> <7> <8> <9> <10>
    <10.95> <12> <14.4> <17.28> <20.74> <24.88>
    mathx10}{}

  \DeclareSymbolFont{mathx}{U}{mathx}{m}{n}

  \DeclareFontFamily{U}{mathb}{\hyphenchar\font45}

  \DeclareFontShape{U}{mathb}{m}{n}{
    <5> <6> <7> <8> <9> <10>
    <10.95> <12> <14.4> <17.28> <20.74> <24.88>
    mathb10}{}

  \DeclareSymbolFont{mathb}{U}{mathb}{m}{n}

  \DeclareMathAccent{\widebar}{0}{mathx}{"73}

  \DeclareMathSymbol{\Rsh}{\mathrel}{mathb}{"E9}

  \DeclareFontFamily{U}{MnSymbolA}{}

  \DeclareFontShape{U}{MnSymbolA}{m}{n}{
    <-6> MnSymbolA5
    <6-7> MnSymbolA6
    <7-8> MnSymbolA7
    <8-9> MnSymbolA8
    <9-10> MnSymbolA9
    <10-12> MnSymbolA10
    <12-> MnSymbolA12}{}

  \DeclareSymbolFont{MnSyA}{U}{MnSymbolA}{m}{n}

  \DeclareMathSymbol{\twoheaddownarrow}{\mathrel}{MnSyA}{27}

  \newcommand{\MSC}[1]{%
    \let\thempfn\relax
    \footnotetext[0]{2020 Mathematics Subject Classification: #1.}
  }

\tikzstyle{vertex}=[circle, draw, minimum size=7pt, inner sep=0pt]

\author{Jacob Ender \and Krzysztof Kapulkin} 
\title{Fast computation of the first discrete homology group}
\date{\today}

\begin{document}

  \maketitle

  \begin{abstract}
    We present a new algorithm for computing the first discrete homology group of a graph.
    By testing the algorithm on different data sets of random graphs, we find that it significantly outperforms other known algorithms.
    \MSC{05-04 (primary), 05C25, 55U15 (secondary)}
  \end{abstract}

\section*{Introduction}

Introduced only a little more than a decade ago \cite{barcelo-capraro-white}, discrete homology has emerged as a crucial graph invariant in a variety of applications.
Part of the broader field of discrete homotopy theory \cite{babson-barcelo-longueville-laubenbacher,barcelo-laubenbacher}, it is closely related via the Hurewicz theorem \cite{carranza-kapulkin:cubical-setting} to discrete homotopy groups, and thus detects important information about the graph from the point of view of matroid theory, hyperplane arrangements, and topological data analysis, among many others \cite{barcelo-laubenbacher,kapuklin-kershaw:data-analysis}.
Consequently, much work has been put into understanding this invariant, including: comparison to other `homology theories' of graphs \cite{barcelo-greene-jarrah-welker:comparison}, results on vanishing \cite{barcelo-greene-jarrah-welker:vanishing}, and theoretical results on ways of speeding up software computations thereof \cite{barcelo-greene-jarrah-welker:connections,greene-welker-wille}.
The first software computations of discrete homology date back to \cite{barcelo-greene-jarrah-welker:comparison} and have since been improved in \cite{kapuklin-kershaw:computations}, showing that one can reasonably compute the third discrete homology group of many graphs encountered in practice.

In the present paper, we focus on computing the first discrete homology group of a graph, following the work of \cite{barcelo-greene-jarrah-welker:comparison,kapuklin-kershaw:computations}, and introduce a new algorithm for doing so.
The algorithm is based on a simple observation that the first discrete homology group of a graph $G$ can be computed instead as the first (cellular) homology group of a cell complex $X_G$ associated to the graph.
The complex $X_G$ has $G$ as its 1-skeleton and 3- and 4-cycles of $G$ as boundaries of its 2-cells.
The boundary matrix of the first two differentials in the cellular chain complex of $X_G$ is typically much smaller than the matrix of the corresponding differentials in the chain complex computing the discrete homology of $G$.
The underlying observation was first made in the context of discrete homotopy groups in \cite{barcelo-kramer-laubenbacher-weaver}, and can be adopted to the setting of discrete homology thanks to the discrete Hurewicz theorem \cite{carranza-kapulkin:cubical-setting,carranza-kapulkin-tonks:hurewicz-cubical}.

Our algorithm consistently outperforms other existing algorithms, including those used in \cite{barcelo-greene-jarrah-welker:comparison,kapuklin-kershaw:computations}.
These previous algorithms are not optimized for computing the \emph{first} homology group, instead focusing on computing homology in as high dimensions as possible with reasonable computational time and resources.
Our focus on the first discrete homology group comes from the fact that it is the one most commonly used in the applications of discrete homotopy theory to topological data analysis \cite{kapuklin-kershaw:data-analysis}.
In particular, our algorithm is compatible with the persistence algorithm \cite{zomorodian-carlsson}.

This begs the question, however, of whether similar improvements can be made in higher homological degrees.
Identifying the 3- and 4-cycles as the unique shapes that need to be glued in dimension 2 does not provide an immediately clear path forward for what such shapes might be in higher dimensions.
More to the point, there are too many possible generalizations of 3- and 4-cycles to higher dimensions that lead to provably wrong conjectures.
The problem of finding a suitable such generalization, termed the \emph{basic shapes conjecture}, remains among the most difficult and important open problems in the field of discrete homotopy theory.

This paper is organized as follows.
In \cref{sec:prelims}, we review the necessary theoretical background on discrete homology, followed by the summary of algorithms in \cref{sec:computations}.
We present our algorithm in \cref{sec:cellular} and, in \cref{sec:comparison}, discuss the comparison of running times of our algorithm in relation to the existing ones.
The pseudocode for our algorithm can be found in \cref{sec:pseudo} and the Julia implementation used for the running times is available at \url{https://github.com/JacobEnder/DiscreteHomology-Algorithms}.

\section{Discrete homology} \label{sec:prelims}

In this section, we review the necessary background on discrete homology of graphs.
Originally introduced in \cite{barcelo-capraro-white} in the context of metric spaces, studied via an associated filtration of graphs, discrete homology was subsequently studied almost exclusively in the graph-theoretic context \cite{barcelo-greene-jarrah-welker:comparison,barcelo-greene-jarrah-welker:vanishing,carranza-kapulkin:cubical-setting,kapuklin-kershaw:computations}.

As our algorithm works best over the field $\mathbb{Z}/2$, we will only consider homology with $\mathbb{Z}/2$-coefficients.
To this end, throughout the paper, all vector spaces will be over $\mathbb{Z}/2$.

We begin by reviewing the definition of a graph and a graph map.
Informally speaking, we are working with simple, undirected, and reflexive graphs and graph homomorphisms; formally:

\begin{definition} \label{def:graph} \leavevmode
\begin{enumerate}
    \item A \emph{graph} $G$ consists of a set $G_V$ together with a reflexive and symmetric relation $\sim \subseteq G_V \times G_V$, called the \emph{adjacency} relation. 
    The set $G_V$ is called the set of \emph{vertices} of $G$, and the relation $\sim$ is the set of \emph{edges} of $G$. 
    We write $v \sim w$ to indicate that $(v,w) \in \sim$.
    \item For graphs $G$ and $H$, a function $f \colon G_V \to H_V$ is a \emph{graph map} if $f$ preserves the adjacency relation.
\end{enumerate}
\end{definition}

\begin{example} \label{def:graph_exs}
The graph $I_n$ has vertex set $\{ 0, 1, \dots, n \}$, with edges $(i, i+1)$ for $0 \leq i \leq n-1$. 

\end{example}

Different notions of graph product have been studied extensively in the literature \cite{imrich-klavzar,kapulkin-kershaw:monoidal-graphs,grenier-kapulkin}, but for the purposes of defining discrete homology, we only require the box product, which we now introduce.

\begin{definition} \label{def:boxprod}
Given graphs $G$ and $H$, their \emph{box product} $G \square H$ has vertex set $G_V \times H_V$, and edges are pairs $((u, v), (u', v'))$, where either $u = u'$ and $v \sim v'$ or $u \sim u'$ and $v = v'$. 
\end{definition}

Discrete homology is the homology of the chain complex of non-degenerate singular $n$-cubes in a graph.
The following definition summarizes the combinatorial aspects of the construction.

\begin{definition} \label{def:cubes}
\begin{enumerate}
    \item\label{def:discrete_n_cube} The \emph{discrete $n$-hypercube} is the graph $I_1^{\square n}$, with vertices are labeled $(x_1, \dots, x_n)$, for $x_i \in \{0,1\}$ for $0 \leq i \leq n$.
    \item\label{def:sing_n_cube} A \emph{singular $n$-cube} in a graph $G$ is a graph map $A \colon I_1^{\square n} \to G$.
    \item\label{def:face_maps}  For a graph $G$, an integer $n \geq 1$, a singular $n$-cube $A \colon I_1^{\square n} \to G$, and $1 \leq i \leq n$, define the \emph{$i$-th positive and negative face} of $A$ as:
    \begin{align*}
        \delta_i^+A(x_1,\dots,x_{n-1}) &:= A(x_1, \dots, x_{n-1}, 1, x_i, \dots, x_n) \\
        \delta_i^-A(x_1,\dots,x_{n-1}) &:= A(x_1, \dots, x_{n-1}, 0, x_i, \dots, x_n). 
    \end{align*}
    \item A singular $n$-cube $A \colon I_1^{\square n} \to G$ is \emph{degenerate} if $\delta_i^+ A = \delta_i^- A$ for some $i$. 
    A singular $n$-cube is non-degenerate if it is not degenerate.
\end{enumerate}
\end{definition}

From this point on, we only need simple algebra:

\begin{definition} \label{def:modules}
    For a graph $G$, define $C_n(G)$ to be the free vector space (over $\mathbb{Z}/2$) on non-degenerate singular $n$-cubes in $G$.
    These assemble into a chain complex with the differential defined on singular $n$-cubes by
     \[
    \partial_n(A) := \sum_{i=1}^n (-1)^i (\delta_i^-A - \delta_i^+A)\text{.} 
    \]
    The \emph{discrete homology} of $G$ is the homology of this chain complex, i.e., $\mathcal{H}_n(G) = \ker(\partial_n) / \operatorname{im}(\partial_{n+1})$.
\end{definition}

Of course, there is some work involved in showing that $C_n$'s with the differential defined above indeed form a chain complex; this was verified in \cite{barcelo-capraro-white}.

\section{Computations} \label{sec:computations}

Following \cite{kapuklin-kershaw:computations}, we now review two existing algorithms for computing discrete homology of a graph, which we call the \emph{cubical algorithm} and the \emph{edge-graph algorithm}, respectively.
For each algorithm, we give a brief description followed by complexity analysis.
When discussing the complexity of the algorithms, we will use $n$ for the number of vertices of $G$ and $m$ for the number of edges.
The first computer-aided computations of discrete homology appear in \cite{barcelo-greene-jarrah-welker:vanishing}, but their algorithm is now superceded by \cite{kapuklin-kershaw:computations}.

\paragraph{The cubical algorithm.}
The cubical algorithm is an algorithm for computing arbitrary homology groups $\mathcal{H}_n(G)$ and as such it is not optimized for efficiency in computing the first homology group specifically. The key insight is the fast generation of singular $(n+1)$-cubes in a graph: instead of considering arbitrary set maps and verifying whether they are valid graph maps, it is observed that singular $n$-cubes in $G$ can be formed inductively by pairing two $(n-1)$-cubes by checking that the corresponding vertices are connected by an edge. The algorithm proceeds as follows. Detailed pseudo-code is available in \cite{kapuklin-kershaw:computations}.
\begin{enumerate}
    \item Generate lists of 0-cubes (vertices) and 1-cubes (edges). This step is $O(1)$.
    \item Identify non-degenerate 1-cubes and construct the matrix of the boundary operator $\partial_1$, then compute $\dim(\ker(\partial_1)) = m - \operatorname{rank}(\partial_1)$. This rank computation is $O(n^4)$ in the worst case, so this entire step is $O(n^4)$. 
    \item For every pair of 1-cubes, check if they form a valid, non-degenerate 2-cube by checking the appropriate adjacencies and ensuring that the 1-cubes do not share a degeneracy direction. This step is $O(n^4)$.  
    \item Identify non-degenerate 2-cubes, construct the matrix of the boundary operator $\partial_2$, and compute $\operatorname{rank}(\partial_2)$. The most expensive part of this step is computing $\operatorname{rank}(\partial_2)$ by Gaussian elimination. The number of rows of $\partial_2$ is order $O(n^2)$, and the number of columns is $O(n^4)$. Thus, the time complexity of this step is $O(n^2 \cdot n^4 \cdot \min(n^2, n^4)) = O(n^8)$.
    \item Return $\dim(\mathcal{H}_1(G)) = \dim(\ker(\partial_1)) - \operatorname{rank}(\partial_2)$. This is $O(1)$.
\end{enumerate}
The entire cubical algorithm has time complexity $O(n^8)$, as it is dominated by the computation of $\operatorname{rank}(\partial_2)$. 

\paragraph{The edge graph algorithm.}
Given a graph $G$, its \emph{edge graph} is the graph $E(G)$ whose vertices are edges in $G$, with vertices connected by an edge if they are opposite edges in a 4-cycle. In dimension 1, we can compute discrete homology without constructing the edge graph explicitly. Instead, it suffices to enumerate a certain subset of the 4-cycles in $G$. To find these 4-cycles, we use a procedure that can be used to construct the edge graph, following step 5 below. Rather than explicitly computing the edge graph, we may use the cycles that we find as the set of 2-cubes, and construct the boundary matrix from these. Pseudocode for the edge graph algorithm can be found in \cref{sec:pseudo}.  An implementation of this method in dimension 1 due to the authors of that paper can be found at \url{https://github.com/JacobEnder/DiscreteHomology-Algorithms/tree/master/edge_graph_homology}. The algorithm functions as follows.
\begin{enumerate}
    \item For each vertex $v \in G_V$, construct its \emph{neighborhood} $N(v)$, the list of all vertices adjacent to $v$. This step is $O(n^2)$. 
    \item Build the chain groups $C_0$ and $C_1$. $C_0$ is just the vertex set $G_V$, and $C_1$ is the edge set $G_E$, excluding self-loops of vertices. This step is $O(1)$. 
    \item Construct the matrix of the boundary operator $\partial_1$, and compute $\dim(\ker(\partial_1))$. This rank computation is again $O(n^4)$ in the worst case.
    \item To assist in building the matrix of $\partial_2$, construct a coordinate dictionary mapping each edge to its column index in $\partial_1$. This step is $O(n^2)$.
    \item Find all 2-cubes by the following procedure. For all $v \in G_V$, get the neighborhood $N(v)$ and consider $w, v' \in N(v)$. If $v' > w > v$ (using the natural ordering of the vertices), then any $w' \in N(w) \cap N(v')$ with $w' \geq v$ forms a non-degenerate 2-cube with vertices $v, w, v', w'$. We also have to consider other cases in which we need to alter these inequalities, to ensure we get all necessary 2-cubes. This process is $O(n^4)$. 
    \item Construct the matrix of $\partial_2$ and compute its rank. Much like the cubical algorithm, the complexity of this rank computation is $O(n^8)$.
    \item Return $\dim(\mathcal{H}_1(G)) = \dim(\ker(\partial_1)) - \operatorname{rank}(\partial_2)$. This is again $O(1)$.
\end{enumerate}
The worst-case time complexity of the edge graph algorithm is $O(n^8)$, the same as the cubical algorithm. However, in practice, the edge graph algorithm is faster on sparser graphs than the cubical algorithm. This is because the cubical algorithm does a naive search over all pairs of edges to search for ways to pair 1-cubes into 2-cubes, but the edge graph method does a local search of the neighborhood of each vertex. In the worst case of a complete graph, these two procedures take the same amount of time, but for denser graphs, the edge graph search is considerably faster.

\section{The cellular algorithm} \label{sec:cellular}

As it turns out, the first discrete homology group admits a topological realization in the sense that there is a topological space $X_G$ one can associate to a graph $G$ with the property that the discrete homology of $G$ coincides with the (singular) homology of $X_G$.
This was originally done \cite[Prop.~5.12]{barcelo-kramer-laubenbacher-weaver} in the context of discrete \emph{homotopy} groups.
However, the first homology group is the abelianization of the first homotopy group, both discretely and classically, making the desired result an immediate consequence.

In order to construct $X_G$, we recall the definition of a simple cycle in a graph.

\begin{definition} \label{def:simple_cycle}
    For $n \geq 3$, a \emph{simple n-cycle} in a graph $G$ is a sequence of distinct vertices $(v_0, \dots, v_{n-1})$ such that $v_i \sim v_{i+1}$ for $0 \leq i \leq n-2$, $v_{n-1} \sim v_0$, and no proper subsequence of $(v_0, \dots, v_{n-1})$ satisfies these conditions.
\end{definition}

\begin{example} \label{ex:simple_cycle}
    Every $3$-cycle in a graph is simple.
    The figure below shows an example of a simple 4-cycle, and an example of a non-simple 4-cycle.
    \[
        \begin{tikzpicture}[every node/.style={circle, fill=black, inner sep=1.5pt}]
            \node (A) at (0,0) {};
            \node (B) at (1,0) {};
            \node (C) at (1,1) {};
            \node (D) at (0,1) {};
        
            \draw (A) -- (B) -- (C) -- (D) -- (A);
        
            \node[rectangle, draw=none,fill=none, outer sep = 0pt, inner sep = 0pt] at (0.5,-0.5) {A simple 4-cycle.};
        
            \node (E) at (4.5,0) {};
            \node (F) at (5.5,0) {};
            \node (G) at (5.5,1) {};
            \node (H) at (4.5,1) {};
        
            \draw (E) -- (F) -- (G) -- (H) -- (E);
            \draw (E) -- (G); 
        
            \node[rectangle, draw=none,fill=none, outer sep = 0pt, inner sep = 0pt] at (5,-0.5) {A non-simple 4-cycle.};
        \end{tikzpicture}
    \]
\end{example}

Having the definition of simple $n$-cycles, we build the desired CW complex as follows.

\begin{definition} \label{def:xg}
    Fix a graph $G$, and define a 2-dimensional CW-complex $X_G$ by declaring its 1-skeleton to be the graph $G$, with a single 2-cell glued into every simple $3$- and $4$-cycle.
\end{definition}

With that, we can now state the theorem underlying our algorithm:

\begin{theorem}\label{thm:hlgy_correspondence}
    For a connected graph $G$, we have $\mathcal{H}_1(G) \iso H_1(X_G)$, where $H_1$ denotes the singular homology of a topological space.
\end{theorem}

As indicated above, the proof depends on the prior results of Barcelo, Kramer, Laubenbacher, and Weaver on discrete homotopy groups.
For brevity, we do not recall them here in detail, instead referring the reader unfamiliar with these notions directly to any of \cite{barcelo-kramer-laubenbacher-weaver,babson-barcelo-longueville-laubenbacher,carranza-kapulkin:cubical-setting}

\begin{proof}
  By \cite[Prop.~5.12]{barcelo-kramer-laubenbacher-weaver} (cf.~\cite{carranza-kapulkin:cubical-setting}), $A_1(G) \iso \pi_1(X_G)$, where $A_1$ denotes the first discrete homotopy group of $G$.
  Using the discrete \cite[Thm.~4.1]{barcelo-capraro-white} and classical \cite[\S2.A]{hatcher} one-dimensional Hurewicz theorems, the abelianization of the left hand side is $\mathcal{H}_1(G)$ and the abelianization of the right hand side is $H_1(X_G)$, thus completing the proof.
\end{proof}

\subsubsection{The cellular algorithm} \label{sssec:cellular_alg}

By \cref{thm:hlgy_correspondence}, we may compute $\mathcal{H}_1(G)$ by computing $H_1(X_G)$ instead, and thus our algorithm first finds 3- and 4-cycles in the graph, then computes the requisite cellular differentials, before finding the rank of the matrix using Gaussian elimination.
With that, we can describe our algorithm as follows:
\begin{enumerate}
    \item\label{cellular-algo:3-and-4-cycles} Find all 3- and 4-cycles in the graph $G$.
    \item\label{cellular-algo:matrix} Create the (sparse) matrix $M$ whose columns are indexed by the edges and 3- and 4-cycles and whose rows are indexed by the edges and vertices of $G$ as follows:
    if a vertex/edge $\sigma_i$ is appears in the edge/cycle $\sigma_j$, we set $M_{ij} = 1$; otherwise, we set $M_{ij} = 0$.
    Thus $M$ represents the first two differentials in the chain complex computing  $X_G$ (with coefficients in $\mathbb{Z}/2$). 
    \item\label{cellular-algo:table} Create a lookup table maintaining the column indices of $M$ corresponding to boundaries of edges and column indices corresponding to boundaries of 3- and 4-cycles. Denote the blocks of $M$ given by these two sets of column indices by $M_1$ and $M_2$, respectively.
    \item\label{cellular-algo:hlgy} Compute $H_1(X_G) = m - \operatorname{rank}(M_1) - \operatorname{rank}(M_2)$. By \cref{thm:hlgy_correspondence}, this is equal to $\mathcal{H}_1(G)$.   
\end{enumerate}
We now discuss each step of the algorithm in detail.

\textbf{\cref{cellular-algo:3-and-4-cycles}.~Finding 3- and 4-cycles.}
The first step of the algorithm enumerates all 3- and 4-cycles in the graph which are the 2-cells of the CW-complex $X_G$.
To list 3-cycles, we use an implementation of the K3 algorithm \cite{k3}, which intersects the neighbourhoods of adjacent vertices, reporting each triangle once.
The time complexity of this approach is at most $O(m^{3/2})$ or $O(n^3)$, since $m = O(n^2)$.
To enumerate 4-cycles, we use a multi-threaded implementation of the 4-cycle listing algorithm of \cite{abboud}, which enumerates all 2-paths in $G$ (that is, triples of vertices $(u, v, w)$ where $u \sim v$ and $v \sim w$), and checks for 2-paths with matching endpoints, indicating a simple 4-cycle in $G$.
The computational complexity of this approach is $O(n^2+t)$, where $t$ is the number of simple 4-cycles in $G$, i.e., $O(n^4)$, so this algorithm is $O(n^4)$. 
Both algorithms work much faster, however, in the typical cases of interest which are generally quite sparse.
Pseudocode for both the K3 algorithm and the 4-cycle enumeration algorithm can be found in their respective papers.

\textbf{\cref{cellular-algo:matrix}.~Building the boundary matrix.}
We build the sparse \emph{boundary matrix} $M$ of the first two differentials in the cellular chain complex of $X_G$.
We work over $\mathbb{Z}/2$, and vertices are stored as integers.
In keeping with the structure of $X_G$, columns of $M$ encode boundaries of edges, and 3- and 4-cycles.
If column $j$ corresponds to an edge, we insert $1$ in column $j$ at the row indices corresponding to its endpoints. If column $j$ corresponds to a 3- or 4-cycles, we insert 1s at the row indices corresponding to its constituent edges. 
The requisite vertex and edge searches (within edges and cycles) require $O(1)$ lookups per cell.
For instance, finding the constituent edges of a 3-cycle requires three lookups, one for each edge. This means that we must perform $O(m+C)$ lookups, where $C$ is the total number of 3- and 4-cycles in $G$.
In the worst case, the number of 4-cycles in $G$ is $O(n^4)$, so building $M$ is also $O(n^4)$.

\textbf{\cref{cellular-algo:table}.~Maintaining a lookup table.}
Once $M$ has been constructed, for convenience, we maintain a lookup table describing which columns correspond to edges and 3- and 4-cycles, so that these do not need to be inferred or re-computed later on.
The columns of $M$ are ordered so that columns corresponding to edges appear before columns corresponding to 3- and 4-cycles. Thus, the column indices corresponding to boundaries of edges are the indices between $n+1$ and $n+m$.
Similarly, columns corresponding to 3- and 4-cycles are found between indices $n+m+1$ and the largest column index in $M$.
To create the lookup table, we need only store some fixed indices in a dictionary, making this step constant $O(1)$. 

\textbf{\cref{cellular-algo:hlgy}.~Computing $\mathcal{H}_1(G).$}
The computation of the first discrete homology group of $X_G$ is a straightforward rank computation.
For a graph $G$, denote the block of $M$ with columns corresponding to the boundaries of edges by $M_1$, and denote the block of $M$ with columns corresponding to boundaries of 3- and 4-cycles by $M_2$.
We calculate $\mathcal{H}_1(G) = m - \operatorname{rank}(M_1) - \operatorname{rank}(M_2)$, using our lookup table to find the necessary ranges of column indices.
Rank computation over $\mathbb{Z}/2$ is done using the \texttt{Modulo2} Julia package, which provides a highly optimized rank computation via Gaussian elimination, using low-level instructions to add columns.
Rank computation is the costliest stage of the cellular algorithm.
Once again letting $C$ denote the number of simple 3- and 4-cycles in $G$, the size of the entire boundary matrix $M$ is $(n + m) \times (m + C)$. In the worst case, $n + m$ is $O(n^2)$. We saw before that in the worst case, $C$ is $O(n^4)$, so $m + C$ is $O(n^4)$. The time complexity of Gaussian elimination on an $k \times \ell$ matrix is $O(k\ell \cdot \min(k, \ell))$.
Thus in the case of a complete graph, rank computation is $O(n^2 \cdot n^4 \cdot n^2) = O(n^8)$.
Pseudocode for the construction of the boundary matrix and for the overall cellular homology computation can be found in \cref{sec:pseudo}.

\section{Comparison of running times} \label{sec:comparison}

We compare the running times of the cellular, cubical and edge graph methods of computing 1-dimensional discrete homology by testing each algorithm on an experimental data set. This data set consists of 800 Erd\H os--R\`enyi graphs (i.e., graphs $G(n,p)$ with $n$ vertices and with any pair of vertices connected by an edge independently with fixed probability $p \in [0,1]$).
The data set was split into four categories:
\begin{enumerate}
    \item 200 graphs with a fixed edge probability $p = 0.07$, and a random number $n$ of vertices, with $100 \leq n \leq 300$;
    \item 200 graphs with a fixed edge probability $p = 0.13$, and a random number $n$ of vertices, with $100 \leq n \leq 300$;
    \item 200 graphs with 100 vertices and random edge probability $p \in [0.07, 0.13]$;
    \item 200 graphs with 300 vertices and random edge probability $p \in [0.07, 0.13]$.
\end{enumerate}
Our  choice of a relatively low probability parameter $p \leq 0.13$ is a consequence of the fact that sparser graphs are more likely to have non-trivial first homology group.
For sufficiently well-connected graphs, the pre-processing techniques of \cite{kapuklin-kershaw:computations} can be applied more effectively.
Here, we wanted to focus on graphs for which pre-processing is unlikely to offer increased efficiency.

The data set was split into these four categories to control any effects of varying the parameters of our Erd\H os--R\`enyi graphs.
The bounds on $n$ and $p$ were chosen to guarantee the presence of both relatively sparse and relatively dense (while still maintaining a good chance of non-trivial first homology) graphs with varying sizes.
After running each of the three algorithms on each of these four sets of graphs, runtime results were collected to form plots for comparison.
All of these plots are available at \url{https://github.com/JacobEnder/DiscreteHomology-Algorithms/tree/master/results}, and we include some of them below.
The first plot shows runtime versus the edge probability $p$ for each of the three algorithms. For the two groups of graphs with fixed edge probability, we display box plots for readability. All times are measured in seconds.

\begin{figure}[H] 
    \centering 
    \includegraphics[scale=0.43]{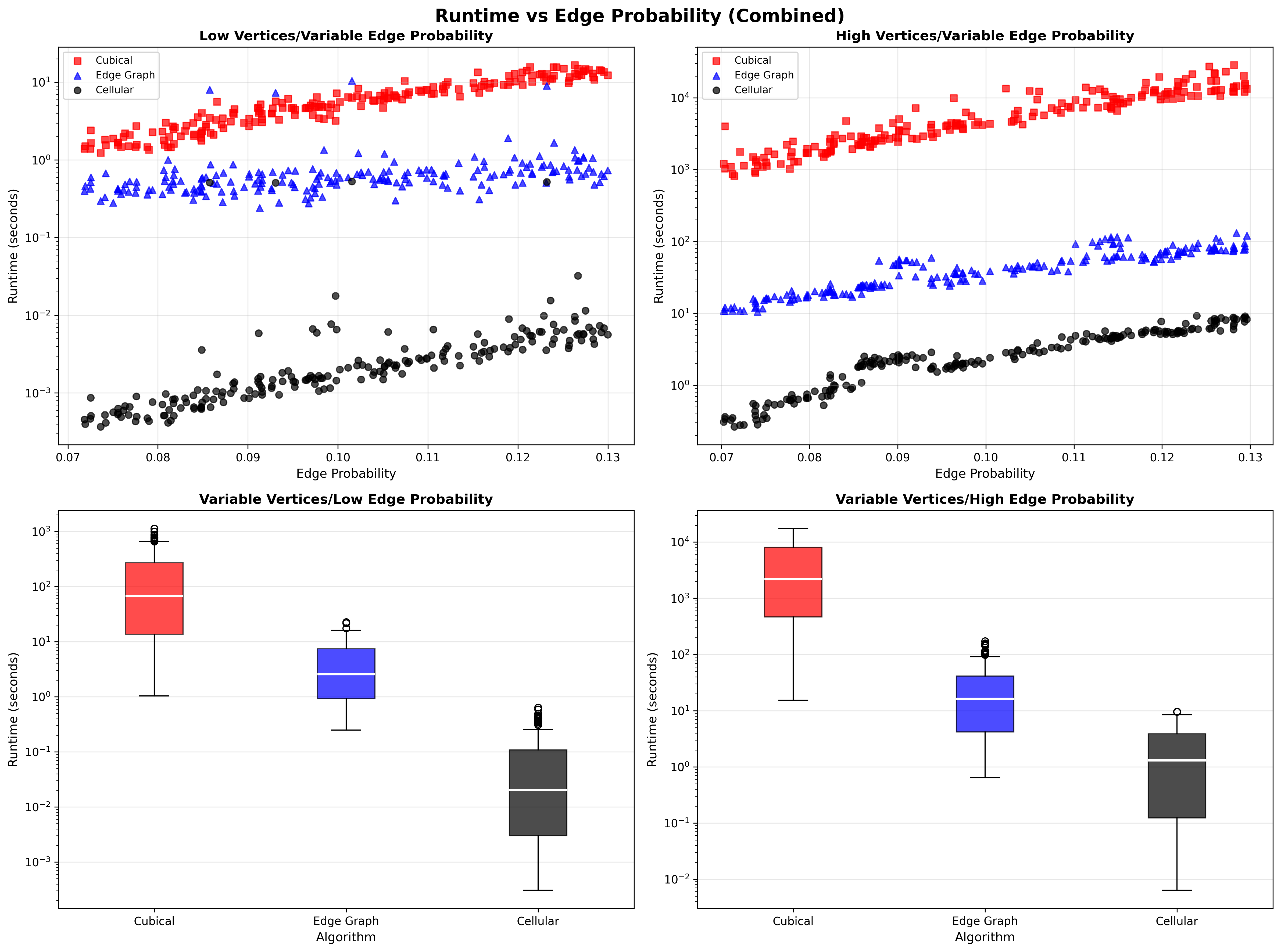} 
    \caption{Runtimes of each of the three algorithms versus edge probability.}
    \label{fig:runtime_plot} 
\end{figure}

One sees a clear separation in the performances of each of the three algorithms.
In this experiment, the cubical algorithm was outperformed by the edge graph algorithm, which was in turn outperformed by the cellular algorithm.
One should observe from the plots that it appears that the runtimes for each algorithm are approximately the same functions of the edge probability $p$, but shifted by some constant factor (this is most apparent in the top-right plot of \cref{fig:runtime_plot}). This is in line with the complexity analyses we performed above.
We found that each algorithm has the same worst-case asymptotic complexity, but there are constant factors that alter computation time in practice.
For instance, the edge graph method must enumerate all 2-cubes in $G$, whereas the cellular algorithm enumerates only simple 3- and 4-cycles.
This makes the boundary matrix smaller in the cellular algorithm, contributing to some practical speedup, especially when $\dim(\mathcal{H}_1(G))$ is high. Moreover, there is fixed computational overhead present in the implementation of the edge graph algorithm that is not present in the cellular algorithm.
This includes dense vector allocation for each 2-cube in $G$, concatenation of dense vectors, and so on.
The cellular algorithm has less of this computational overhead, leading to significant practical performance improvements.
It is possible that there are optimizations that can be applied to the current implementation of the edge graph method that could improve runtimes, but the authors are not aware of such optimizations at present.
The same sort of analysis reveals why the cubical algorithm is even slower than the edge graph algorithm in practice.

Many more plots of these results are available in the GitHub repository found at \url{https://github.com/JacobEnder/DiscreteHomology-Algorithms/tree/master/results}. These plots show pairwise comparisons of the cubical, edge graph, and cellular algorithms against $\dim \mathcal{H}_1(G)$, the edge probability $p$ of the Erd\H os--R\`enyi graphs, the quantity $n$, and the total number of simple 3- and 4-cycles in the graph. Also contained in the \href{https://github.com/JacobEnder/DiscreteHomology-Algorithms/tree/master/results}{results} directory is a file \href{https://github.com/JacobEnder/DiscreteHomology-Algorithms/blob/master/results/detailed_results.csv}{detailed\_results.csv} that shows the following data for each graph $G$ in our experimental data set: graph name, number of vertices, edge probability, number of simple 3-cycles, number of simple 4-cycles, the sum of these quantities, $\dim \mathcal{H}_1(G)$, runtimes of each of the three algorithms on $G$, pairwise ratios of each of the three runtimes, and which algorithm was fastest.

The cellular algorithm described in this paper was the fastest method for every graph in the data set, so the last column is constant.

\appendix
 \renewcommand{\thesection}{\Alph{section}}
 \begin{appendices}
   \section{Pseudocode for some algorithms} \label{sec:pseudo}

Throughout the section, $\operatorname{rank}_{\mathbb{Z}/2}$ refers to the specialized rank function in the \texttt{Modulo2} package.

\begin{algorithm}[H]
\caption{Edge Graph Method for Computing $\mathcal{H}_1(G)$}
\begin{algorithmic}[1]

\STATE \textbf{Input:} Graph $G$ with neighborhood dictionaries
\STATE \textbf{Output:} $\dim \mathcal{H}_1(G)$

\medskip
\STATE \textbf{Step 1: Construct 1-chains and 1-boundary matrix}
\STATE $\mathcal{C}_1 \gets \{\{u, v\} : (u,v) \in E, \, u \neq v\}$ \COMMENT{Non-loop edges}
\STATE Initialize $M_1 \in M_{{|V| \times |\mathcal{C}_1|}}(\mathbb{Z}/2)$ as zero matrix
\FOR{each edge $e = \{u, v\} \in \mathcal{C}_1$}
    \STATE $M_1[u, e] \gets 1$
    \STATE $M_1[v, e] \gets 1$
\ENDFOR

\medskip
\STATE \textbf{Step 2: Enumerate 2-chains (square maps)}
\STATE $\mathcal{C}_2 \gets \emptyset$
\FOR{each vertex $v \in V$}
    \FOR{each $w \in N(v)$ with $w > v$}
        \FOR{each $v' \in N(v)$ with $v' > w$}
            \FOR{each $w' \in N(w) \cap N(v')$ with $w' \geq v$}
                \STATE \COMMENT{Found square: $v \sim w$, $v \sim v'$, $w \sim w'$, $v' \sim w'$}
                \STATE Add squares $(v, w, v', w')$ and $(v, v', w, w')$ to $\mathcal{C}_2$
            \ENDFOR
        \ENDFOR
        \STATE Add degenerate squares
    \ENDFOR
\ENDFOR

\medskip
\STATE \textbf{Step 3: Construct 2-boundary matrix}
\STATE Initialize $M_2 \in M_{|\mathcal{C}_1| \times |\mathcal{C}_2|}(\mathbb{Z}/2)$ as zero matrix
\FOR{each square $\sigma = (v, w, v', w') \in \mathcal{C}_2$}
    \STATE $\text{faces} \gets \big[\{v, w\}, \{v', w'\}, \{w, w'\}, \{v, v'\}\big]$
    \FOR{each face $f \in \text{faces}$}
        \IF{$f \in \mathcal{C}_1$}
            \STATE $M_2[\text{index}(f), \text{index}(\sigma)] \mathrel{+}= 1 \pmod 2$
        \ENDIF
    \ENDFOR
\ENDFOR

\medskip
\STATE \textbf{Step 4: Compute homology dimensions}
\STATE $r_1 \gets \text{rank}_{\mathbb{Z}/2}(M_1)$
\STATE $r_2 \gets \text{rank}_{\mathbb{Z}/2}(M_2)$
\STATE $\dim \mathcal{H}_1 \gets |\mathcal{C}_1| - r_1 - r_2$

\medskip
\RETURN $\dim \mathcal{H}_1$

\end{algorithmic}
\end{algorithm}

\begin{algorithm}[H]
\caption{Cellular Method for Computing $\mathcal{H}_1(G)$}
\begin{algorithmic}[1]

\STATE \textbf{Input:} Graph $G$ with adjacency list representation
\STATE \textbf{Output:} $\dim \mathcal{H}_1(G)$

\medskip
\STATE \textbf{Step 1: Extract graph structure}
\STATE $\mathcal{E} \gets \textsc{Edges}(G)$ 
\STATE $\mathcal{T} \gets \textsc{EnumerateTriangles}(G)$ \COMMENT{K3 algorithm}
\STATE $\mathcal{C} \gets \textsc{EnumerateFourCycles}(G)$ \COMMENT{Algorithm due to Abboud et. al}
\STATE $\mathcal{S} \gets \mathcal{T} \cup \mathcal{C}$ \COMMENT{All short cycles}

\medskip
\STATE \textbf{Step 2: Construct boundary matrix}
\STATE $M \gets \textsc{BuildBoundaryMatrix}(V, \mathcal{E}, \mathcal{S})$

\medskip
\STATE \textbf{Step 3: Extract submatrices}
\STATE $M_1 \gets M[\text{vertex rows}, \text{edge columns}]$ \COMMENT{Size $|V| \times |\mathcal{E}|$}
\STATE $M_2 \gets M[\text{edge rows}, \text{cycle columns}]$ \COMMENT{Size $|\mathcal{E}| \times |\mathcal{S}|$}

\medskip
\STATE \textbf{Step 4: Compute homology dimensions}
\STATE $r_1 \gets \text{rank}_{\mathbb{Z}/2}(M_1)$
\STATE $r_2 \gets \text{rank}_{\mathbb{Z}/2}(M_2)$
\STATE $\dim \mathcal{H}_1 \gets |\mathcal{E}| - r_1 - r_2$

\medskip
\RETURN $\dim \mathcal{H}_1$

\end{algorithmic}
\end{algorithm}

\begin{algorithm}[H]
\caption{\textsc{BuildBoundaryMatrix}: Construct Sparse Boundary Operator}
\begin{algorithmic}[1]

\STATE \textbf{Input:} Vertex set $V$, edge set $\mathcal{E}$, short cycles $\mathcal{S} = \mathcal{T} \cup \mathcal{C}$
\STATE \textbf{Output:} Sparse boundary matrix $M \in M_{(|V| + |\mathcal{E}|) \times (|V| + |\mathcal{E}| + |\mathcal{S}|)}(\mathbb{Z}/2)$

\STATE Build edge index map: $\text{idx}[e] \gets$ position of $e$ in $\mathcal{E}$
\STATE Initialize coordinate lists: $I \gets [], \; J \gets [], \; X \gets []$

\medskip
\STATE \textbf{Edge columns: $\partial(\text{edge}) = \text{boundary vertices}$}
\FOR{$k = 1$ to $|\mathcal{E}|$}
    \STATE $(u, v) \gets \mathcal{E}[k]$
    \STATE $\text{col} \gets |V| + k$
    \STATE Append $(u, \text{col}, 1)$ to $(I, J, X)$
    \STATE Append $(v, \text{col}, 1)$ to $(I, J, X)$
\ENDFOR

\medskip
\STATE \textbf{Short cycle columns: $\partial(\text{cycle}) = \text{boundary edges}$}
\FOR{$k = 1$ to $|\mathcal{S}|$}
    \STATE $\sigma \gets \mathcal{S}[k]$
    \STATE $\text{col} \gets |V| + |\mathcal{E}| + k$
    \STATE $\text{edges} \gets \text{Constituent edges of } \sigma$
    \FOR{each edge $e \in \text{edges}$}
        \STATE $\text{row} \gets |V| + \text{idx}[e]$
        \STATE Append $(\text{row}, \text{col}, 1)$ to $(I, J, X)$
    \ENDFOR
\ENDFOR

\medskip
\STATE $M \gets \textsc{SparseMatrix}(I, J, X, \; |V| + |\mathcal{E}|, \; |V| + |\mathcal{E}| + |\mathcal{S}|)$
\RETURN $M$

\end{algorithmic}
\end{algorithm}

 \end{appendices}

 \bibliographystyle{amsalphaurlmod}
 \bibliography{all-refs.bib}

\begin{bibdiv}
\begin{biblist}

\bib{abboud}{incollection}{
      author={Abboud, Amir},
      author={Khoury, Seri},
      author={Leibowitz, Oree},
      author={Safier, Ron},
       title={Listing 4-cycles},
        date={2023},
   booktitle={43rd {IARCS} {A}nnual {C}onference on {F}oundations of {S}oftware
  {T}echnology and {T}heoretical {C}omputer {S}cience},
      series={LIPIcs. Leibniz Int. Proc. Inform.},
      volume={284},
   publisher={Schloss Dagstuhl. Leibniz-Zent. Inform., Wadern},
       pages={Art. No. 25, 16},
         url={https://doi.org/10.4230/lipics.fsttcs.2023.25},
      review={\MR{4695060}},
}

\bib{babson-barcelo-longueville-laubenbacher}{article}{
      author={Babson, Eric},
      author={Barcelo, H\'{e}l\`ene},
      author={de~Longueville, Mark},
      author={Laubenbacher, Reinhard},
       title={Homotopy theory of graphs},
        date={2006},
        ISSN={0925-9899},
     journal={J. Algebraic Combin.},
      volume={24},
      number={1},
       pages={31\ndash 44},
}

\bib{barcelo-capraro-white}{article}{
      author={Barcelo, H\'{e}l\`ene},
      author={Capraro, Valerio},
      author={White, Jacob~A.},
       title={Discrete homology theory for metric spaces},
        date={2014},
        ISSN={0024-6093},
     journal={Bull. Lond. Math. Soc.},
      volume={46},
      number={5},
       pages={889\ndash 905},
}

\bib{barcelo-greene-jarrah-welker:comparison}{article}{
      author={Barcelo, H\'{e}l\`ene},
      author={Greene, Curtis},
      author={Jarrah, Abdul~Salam},
      author={Welker, Volkmar},
       title={Discrete cubical and path homologies of graphs},
        date={2019},
     journal={Algebr. Comb.},
      volume={2},
      number={3},
       pages={417\ndash 437},
}

\bib{barcelo-greene-jarrah-welker:connections}{article}{
      author={Barcelo, H\'{e}l\`ene},
      author={Greene, Curtis},
      author={Jarrah, Abdul~Salam},
      author={Welker, Volkmar},
       title={Homology groups of cubical sets with connections},
        date={2021},
        ISSN={0927-2852},
     journal={Appl. Categ. Structures},
      volume={29},
      number={3},
       pages={415\ndash 429},
}

\bib{barcelo-greene-jarrah-welker:vanishing}{article}{
      author={Barcelo, H\'{e}l\`ene},
      author={Greene, Curtis},
      author={Jarrah, Abdul~Salam},
      author={Welker, Volkmar},
       title={On the vanishing of discrete singular cubical homology for
  graphs},
        date={2021},
        ISSN={0895-4801},
     journal={SIAM J. Discrete Math.},
      volume={35},
      number={1},
       pages={35\ndash 54},
         url={https://doi.org/10.1137/20M1338484},
      review={\MR{4196413}},
}

\bib{barcelo-kramer-laubenbacher-weaver}{article}{
      author={Barcelo, H\'{e}l\`ene},
      author={Kramer, Xenia},
      author={Laubenbacher, Reinhard},
      author={Weaver, Christopher},
       title={Foundations of a connectivity theory for simplicial complexes},
        date={2001},
        ISSN={0925-9899},
     journal={Adv. Appl. Math.},
      volume={26},
      number={1},
       pages={97\ndash 128},
}

\bib{barcelo-laubenbacher}{article}{
      author={Barcelo, H\'{e}l\`ene},
      author={Laubenbacher, Reinhard},
       title={Perspectives on {$A$}-homotopy theory and its applications},
        date={2005},
        ISSN={0012-365X},
     journal={Discrete Math.},
      volume={298},
      number={1-3},
       pages={39\ndash 61},
}

\bib{carranza-kapulkin:cubical-setting}{article}{
      author={Carranza, D.},
      author={Kapulkin, K.},
       title={Cubical setting for discrete homotopy theory, revisited},
        date={2024},
        ISSN={0010-437X},
     journal={Compos. Math.},
      volume={160},
      number={12},
       pages={2856\ndash 2903},
         url={https://doi.org/10.1112/S0010437X24007486},
      review={\MR{4881411}},
}

\bib{carranza-kapulkin-tonks:hurewicz-cubical}{article}{
      author={Carranza, Daniel},
      author={Kapulkin, Krzysztof},
      author={Tonks, Andrew},
       title={The {H}urewicz theorem for cubical homology},
        date={2023},
        ISSN={0025-5874,1432-1823},
     journal={Math. Z.},
      volume={305},
      number={4},
       pages={Paper No. 61, 20},
         url={https://doi.org/10.1007/s00209-023-03352-0},
      review={\MR{4661259}},
}

\bib{k3}{article}{
      author={Chiba, Norishige},
      author={Nishizeki, Takao},
       title={Arboricity and subgraph listing algorithms},
        date={1985},
        ISSN={0097-5397},
     journal={SIAM J. Comput.},
      volume={14},
      number={1},
       pages={210\ndash 223},
         url={https://doi.org/10.1137/0214017},
      review={\MR{774940}},
}

\bib{grenier-kapulkin}{unpublished}{
      author={Grenier, Adrien},
      author={Kapulkin, Krzysztof},
       title={Biclosed monoidal structures on the categories of digraphs and
  graphs},
        date={2025},
        note={Preprint},
}

\bib{greene-welker-wille}{unpublished}{
      author={Greene, Curtis},
      author={Welker, Volkmar},
      author={Wille, Georg},
       title={On the homology of simplicial and cubical sets with symmetries},
        date={2025},
        note={Preprint},
}

\bib{hatcher}{book}{
      author={Hatcher, Allen},
       title={Algebraic topology},
   publisher={Cambridge University Press, Cambridge},
        date={2002},
        ISBN={0-521-79160-X; 0-521-79540-0},
      review={\MR{1867354}},
}

\bib{imrich-klavzar}{book}{
      author={Imrich, Wilfried},
      author={Klav\v{z}ar, Sandi},
       title={Product graphs},
      series={Wiley-Interscience Series in Discrete Mathematics and
  Optimization},
   publisher={Wiley-Interscience, New York},
        date={2000},
        ISBN={0-471-37039-8},
        note={Structure and recognition, With a foreword by Peter Winkler},
      review={\MR{1788124}},
}

\bib{kapulkin-kershaw:monoidal-graphs}{article}{
      author={Kapulkin, Krzysztof},
      author={Kershaw, Nathan},
       title={Closed symmetric monoidal structures on the category of graphs},
        date={2024},
     journal={Theory Appl. Categ.},
      volume={41},
       pages={Paper No. 23, 760\ndash 784},
      review={\MR{4774716}},
}

\bib{kapuklin-kershaw:computations}{unpublished}{
      author={Kapulkin, Krzysztof},
      author={Kershaw, Nathan},
       title={Efficient computations of discrete cubical homology},
        date={2024},
         url={https://arxiv.org/abs/2410.09939},
}

\bib{kapuklin-kershaw:data-analysis}{unpublished}{
      author={Kapulkin, Krzysztof},
      author={Kershaw, Nathan},
       title={Data analysis using discrete cubical homology},
        date={2025},
         url={https://arxiv.org/abs/2506.15020},
}

\bib{zomorodian-carlsson}{article}{
      author={Zomorodian, Afra},
      author={Carlsson, Gunnar},
       title={Computing persistent homology},
        date={2005},
        ISSN={0179-5376,1432-0444},
     journal={Discrete Comput. Geom.},
      volume={33},
      number={2},
       pages={249\ndash 274},
         url={https://doi.org/10.1007/s00454-004-1146-y},
      review={\MR{2121296}},
}

\end{biblist}
\end{bibdiv}

\end{document}